# High quantum efficiency S-20 photocathodes in photon counting detectors

D. A. Orlov[a,*], J. DeFazio[b], S. Duarte Pinto[a], R. Glazenborg[a] and E. Kernen[a]

[a] *PHOTONIS Netherlands BV,*
  *Dwazziewegen 2, 9301 ZR Roden, The Netherlands*
[b] *PHOTONIS USA Pennsylvania,*
  *1000 New Holland Ave., Lancaster, PA 17601, USA*
  *E-mail*: D.Orlov@nl.photonis.com

ABSTRACT: Based on conventional S-20 processes, a new series of high quantum efficiency (QE) photocathodes has been developed that can be specifically tuned for use in the ultraviolet, blue or green regions of the spectrum. The QE values exceed 30% at maximum response, and the dark count rate is found to be as low as 30 Hz/cm$^2$ at room temperature. This combination of properties along with a fast temporal response makes these photocathodes ideal for application in photon counting detectors, which is demonstrated with an MCP photomultiplier tube for single and multi-photoelectron detection.

# Contents

## 1. Introduction

Detection of extremely low intensity light signals (down to a single photon) is a requirement for many scientific and medical applications. PHOTONIS has developed various types of photon counting detectors, including Planacon™, Multi-MCP photomultiplier tubes (MCP-PMT) and Hybrid Photo Diodes (HPD) [1]. These devices all involve vacuum photoemission from a sensitive photocathode, which is a subject of continual development efforts.

The photocathode can be considered as the first stage of the amplifier and thus its properties, such as QE, dark count rate, and response time, are extremely important for the total performance of the detector. The photon detection probability is mainly defined by the QE, while the noise contribution is usually dominated by thermionic emission from the photocathode. Indeed these thermionic electrons are amplified in an identical manner as the photoelectrons, which often makes it impossible to separate dark counts from single photon events. Additionally, the traveling time of the photoelectron within the photocathode before escaping into the vacuum can in some cases limit timing jitter of a device [2]. Measurements done earlier with a dual MCP-PMT with a conventional S-20 photocathode show a single photon time resolution of 34 ps, attributed to the transfer time spread of the multiplying charges through the MCP [3].

The new photocathodes presented here are based on conventional S-20 (alkali antimonide) processes and as such are directly applicable to scientific image intensifiers and MCP-PMTs, but also to large-areas such as in the multi-anode Planacon™ type detectors and HPDs. Below, the new high quantum efficiency (Hi-QE) photocathode properties are discussed and the overall performance is demonstrated for single and multi-photoelectron detection using these new photocathodes in a MCP-PMT.

## 2. Results and discussions

In this section are the important typical data for photocathode performance: the spectrum of quantum efficiency for the UV, Blue and Green Hi-QE transmission mode photocathode series, the evolution of dark count rate, and the Pulse Height Distribution obtained with dual MCP-PMT for single and multi-photoelectron detection.

**2.1 Quantum Efficiency Spectra**

Figure 1 shows the QE spectra for a conventional transmission mode S-20 and the Hi-QE series photocathodes over the 200-700 nm spectral range. For a conventional S-20 process the peak QE of about 25% is reached at 270 nm. The spectral sensitivity range of these photocathodes is wide, with a QE of still about 3-4% at 700 nm. One compromise for this wide sensitivity range is a lower QE at the peak and a higher dark count rate. The latter is a direct consequence of the photocathode deposition process optimized to get higher red-sensitivity. Such photocathodes, as well as new developed Hi-QE photocathodes, can be provided for fast gating applications with a mesh underlay, which allows for a fast gating time down to below 1 ns [4], but it simultaneously reduces average QE by about 10-15%.

The new photocathodes were developed to target specific spectral ranges which are here called Hi-QE UV, Hi-QE Blue and Hi-QE Green. All were grown on fused silica windows, which have a short-wavelength cutoff at approximately 170 nm. These photocathodes exhibit much higher QE values than the conventional S-20, clearly exceeding 30% as seen in Figure 1.

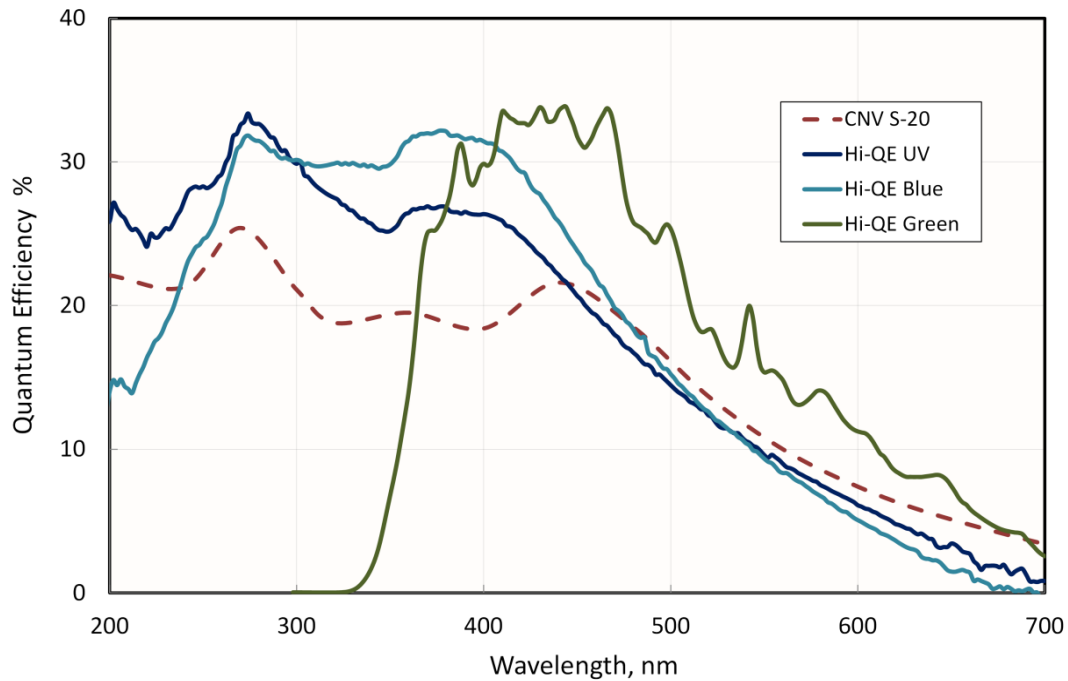

**Figure 1.** Quantum efficiency spectra for newly developed Hi-QE photocathodes: Hi-QE UV (dark blue), Hi-QE Blue (light blue) and Hi-QE Green (green) in comparison with a conventional S-20 photocathode (CNV S-20, dashed red).

Hi-QE UV photocathodes are optimized for the UV range with a maximum QE at 270 nm of typically 31-34%. These photocathodes can also be grown on specially prepared sapphire cathode substrates allowing extension of the sensitivity spectral range down to 150 nm (not shown here).

Hi-QE Blue photocathodes were designed to provide highest QE in the 260-410 nm spectral range. The QE spectrum shows a broad plateau in this range, again with a typical QE above 30%. The decrease of QE below 260 nm (compared to Hi-QE UV) is the trade-off for high sensitivity in blue spectral range.

Hi-QE Green photocathodes demonstrate their QE values above 30% in the range of 390-480 nm, while at 500 nm the QE is still about 25%. Comparing this to the other Hi-QE photocathodes, the sensitivity of Hi-QE Green is much higher at longer wavelengths up to 700 nm. It is important to mention that despite the high sensitivity at long wavelengths, the dark count rate of these photocathodes stays extremely low, typically below 50 Hz/cm$^2$ (as low as the other Hi-QE cathodes), making Hi-QE Green unique for photon counting in this spectral range.

**2.2 Dark Count Rate of Hi-QE Photocathodes**

For low-rate single photon detection, it can be extremely important to minimize the dark count rate as much as possible. Figure 2 shows evolution of the dark count rate with time for MCP-PMTs with conventional S-20 and the newly developed Hi-QE Blue S20 photocathode at room temperature (23°C). For these measurements the detectors were placed in dark conditions at time zero and the MCP-PMT voltages were set to obtain a gain around 10$^5$ (see following section). Count rate was then measured versus time intermittently throughout the next ~ 200 minutes.

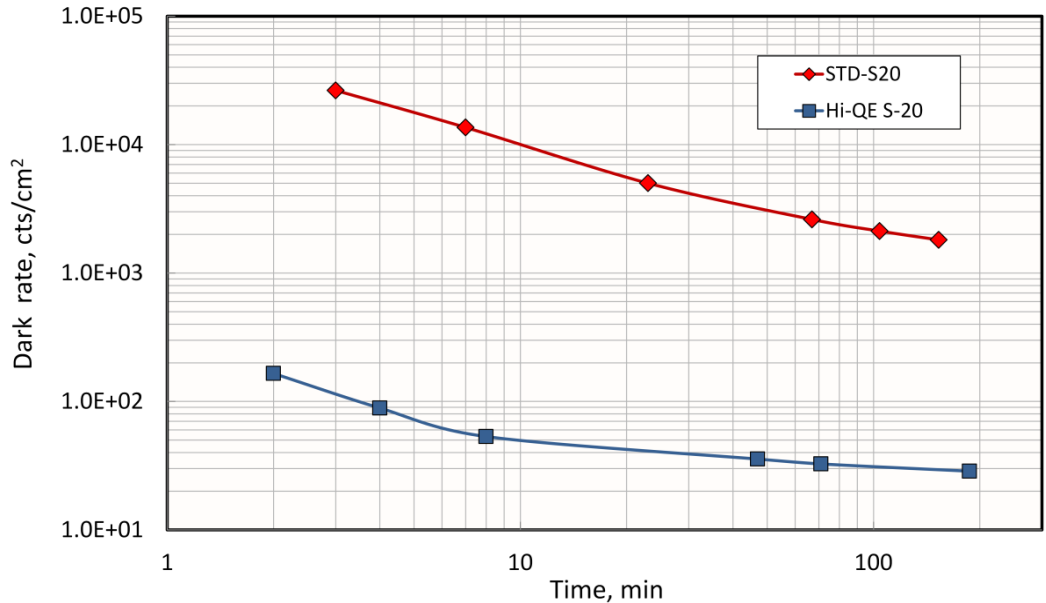

**Figure 2.** Evolution of dark count rate versus time at ~23°C for a conventional S-20 (red) and the new Hi-QE Blue photocathodes (blue). Very low dark count rate < 30 Hz/cm$^2$ can be achieved with the Hi-QE series photocathodes.

One can see in Figure 2 that it takes 2-3 hours to reach the low-rate plateau. The high dark count rate measured initially is hypothesised to originate from long-lived surface and bulk states lying above Fermi level populated by ambient light. It takes time to discharge these states, where the decay time is an important criterion of the detector performance. For Hi-QE photocathodes we adjusted the deposition processes to keep the dark count rate low and have a quick discharge of surface and bulk states.

While in the case of conventional broad-range S20 photocathodes the dark rate approaches 1000-2000 Hz/cm$^2$, the dark rate for Hi-QE photocathodes typically is only about 20-30 Hz/cm$^2$. In the presented case after 10 min in the dark, the dark count rate already drops below 50 Hz/cm$^2$. Tests indicate that all Hi-QE S20 series photocathodes (UV, Blue, Green) behave similarly and exhibit nearly the same values of dark count rate.

## 2.3 Single/Multi Photon PHD Measurements with Hi-QE dual MCP-PMT

Figure 3 presents a Pulse Height Distribution (PHD) obtained using a dual MCP-PMT with Hi-QE photocathode, illustrating the ability for single photoelectron detection. The PHD was recorded using a Charge Sensitive Preamplifier CSP10[1] (1.4 V/pC), a shaping amplifier CSA4[1] (gain=10, shaping time of 250 ns), and a multi-channel analyser MCA3[1] (scale=0.89 mV/ch). The threshold was set at 32 ch.

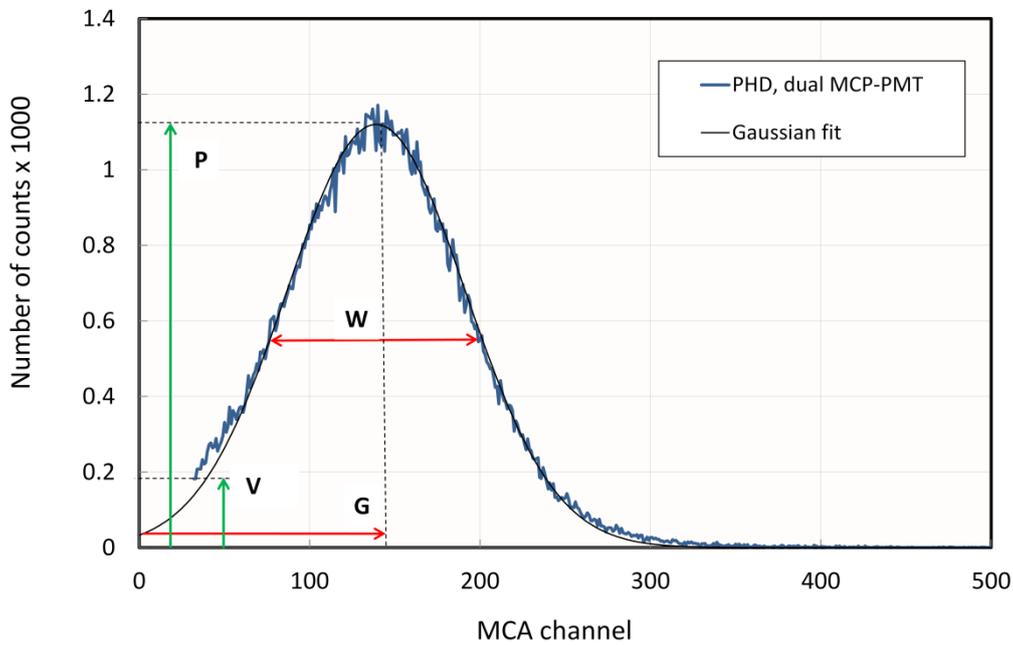

**Figure 3.** Pulse Height Distribution recorded with PHOTONIS dual MCP-PMT and Hi-QE photocathode with single photon illumination (blue). The Gaussian curve (black) is a fit to experimental results.

The PHD shown in Figure 3 was measured with low input background light illumination, keeping the count rate below a few hundred Hz. The MCP voltage of 1625 V (for dual MCP) yields electron gain of 1.1x10$^5$. The dark count rate originating from the photocathode was approximately 30 Hz/cm$^2$, while the MCP contribution was negligible at < 0.2 Hz.

In the resulting PHD (blue curve), the peak is well separated, with a low-energy valley and noise below the threshold. The peak is described well by a Gaussian distribution (solid black curve). The gain point ("G") corresponds to the mean energy of the PHD and is just slightly above the position of the PHD peak. Both the peak-to-valley (P/V) ratio and the full width half maximum-to-gain (W/G) ratio are typically used to characterize photon counting detectors. Here the measured values are P/V≈6 and W/G≈0.86 which are according to our knowledge, the best available for dual MCP PMTs.

---

[1] FAST ComTec Communication Technology GmbH.

Another example demonstrating performance of a PHOTONIS MCP-PMT with Hi-QE photocathode is presented in Figure 4. In this case the photocathode was illuminated with a 100 kHz short-pulse (100 ps) defocused laser beam. The laser intensity was strongly attenuated to yield a few photoelectrons per pulse.

These few-photoelectron measurements were performed using the same detector and settings as in Figure 3. The blue curve in Figure 4 is the measured PHD and the black curve is a fit, which is a sum of seven Gaussian curves. The position ($G_1$) and width ($W_1$) of the first fitting Gaussian curve, corresponding to single-photon illumination, were taken from Figure 3. Other Gaussian curves (Figure 4) correspond to amplification of 2-7 photoelectrons. The positions of these curves were fixed as $G_N=G_1\times N$ and the width was scaled according to statistical rule as $W_N=W_1\times\sqrt{N}$, thus the only fitting parameters were the amplitudes of the Gaussian peaks. A very good fit to the measured PHD is obtained, with clear separation of the first and second peaks corresponding to single and two-electron photoemission. A shoulder for 3 photoelectron emission can also be seen while further peaks are not well resolved due to the statistical increase of the PHD width with the increasing number of photoelectrons.

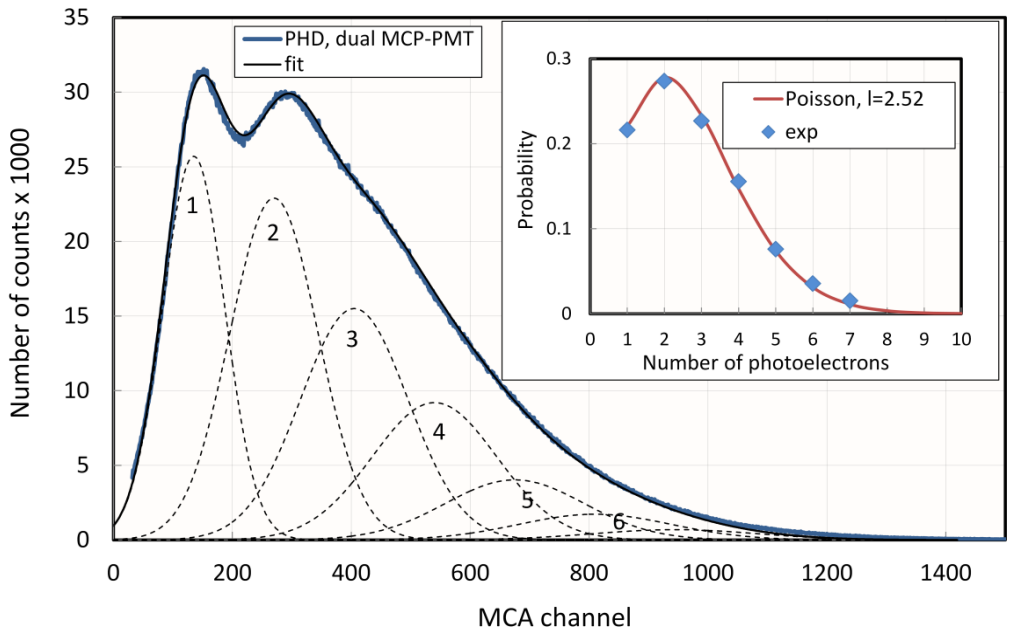

**Figure 4.** Few photoelectron Pulse Height Distribution recorded with a PHOTONIS dual MCP-PMT and Hi-QE photocathode (the same detector and settings as in Figure 3). The fitting curve is a sum (1-7) of Gaussian curves, corresponding to multi (1-7) photoelectron amplification.
**Inset:** the normalized area of each (1-7) Gaussian curve (blue dots). The red curve is a calculated Poisson distribution with expected value $\lambda\approx2.5$ for the average number of electrons emitted from the photocathode.

The intensity (corresponding to the area) of each peak is shown in the inset in Figure 4 (blue points). The red curve is a fit by Poisson distribution with expected value λ of 2.52, therefore this excellent fit clearly indicates the average number of photoelectrons.

The data presented in Figure 4 show that MCP-PMTs are ideal for single photon detection, but also demonstrate their limitations for applications where few photoelectron resolution may be required. The gain difference between "N" and "N+1" photoelectrons is still equal to $G_1$, however the $N^{th}$ peaks become broader, so that effective resolution degrades as $\sqrt{N}$. In comparison, the Hybrid Photo Diodes [5], with a very high internal gain for first/single hit, demonstrate much smaller W/G ratio of about 0.14. In this case even for detection of multiple photoelectrons (~10) all individual peaks are well separated with $W_{10}/G_1$ below 0.5, which is even better than single photoelectron resolution with dual MCP-PMTs.

## 3. Conclusions

High quantum efficiency UV, blue and green transmission mode photocathodes were developed with QEs above 30% achievable in tuneable spectral ranges. Together with an extremely low dark count rate down to ~30 Hz/cm$^2$ and a fast temporal response, these Hi-QE photocathodes are ideal for photon counting devices. The measurements of PHD with single and multiple photoelectrons demonstrate a high performance when integrated into PHOTONIS photon counting detectors.

## 4. Acknowledgements

We would like to thank Teun R., Jack S., Henk V., John vd W., Menno R., Jan H. for useful discussions and for their help optimizing processes to develop these Hi-QE photocathodes.

## References

[1] See https://www.photonis.com/en/photon-detectors

[2] W.E. Spicer and A. Herrera-Gomez, Modern theory and applications of photocathodes, Proceedings of SPIE **2022**, (1993) 18.

[3] L. Castillo Garcia, Systematic studies of microchannel plate tubes model PP0365G from Photonis, LHCb-PUB-2013-017.

[4] W. Uhring et al., 200 ps FWHM and 100 MHz repetition rate ultrafast gated camera for optical medical functional imaging, Proceedings of SPIE **8439** (2012) 84392L.

[5] R. DeSalvo, Why people like the Hybrid PhotoDiode, Nucl. Instrum. Meth. A **387** (1997) 92.